# Plasmon Induced Delocalized Second-Harmonic Generation Towards Buried-Interface Spectroscopy

**Alan R. Bowman[1,2], Sergejs Boroviks[3], Omer Can Karaman[1], Laura M. Herz[2], Olivier J.F. Martin[3] and Giulia Tagliabue[1+]**

1. Laboratory of Nanoscience for Energy Technologies (LNET), STI, École Polytechnique Fédérale de Lausanne (EPFL), Lausanne 1015, Switzerland
2. Clarendon Laboratory, Department of Physics, University of Oxford, Parks Road, Oxford, OX1 3PU, UK
3. Nanophotonics and Metrology Laboratory (NAM), Swiss Federal Institute of Technology Lausanne (EPFL), 1015 Lausanne, Switzerland

+corresponding author: giulia.tagliabue@epfl.ch

## Abstract

Second-harmonic generation microscopy is a powerful technique capable of probing local crystal symmetries and electric fields at interfaces. However, it often suffers from weak signal strength and is difficult to understand in multilayer systems where many materials can give competing signal contributions. In this work we present direct observation of delocalized, surface plasmon polariton-mediated second-harmonic generation on gold monocrystalline surfaces and structures. We generate second-harmonic light up to 35 µm from the excitation spot and, excitingly, we obtain signal from atomically flat surfaces without a fundamental excitation beam present in the same region. We reveal that this process arises from the interaction of two counter-propagating surface plasmon polaritons, which we believe to be the first observation of this process at the microscale. This signal has the same polarisation dependence as localised second-harmonic generation and is emitted in a collimated beam travelling perpendicular to the sample surface. In part due to local electric field enhancements, we were able to observe these signals on a CMOS camera with 1 s exposure and no gain using an industrial-grade pulsed laser. Our results enable wide area multilayer samples to be probed using a single excitation beam, with applications including in energy, catalysis and single particle surface sensing.

## Introduction

Second-harmonic generation (SHG), two photons of angular frequency $\omega$ generating a single photon with frequency $2\omega$, is a nonlinear optical process that was first observed from a bulk nonlinear crystal in 1961 [1] and a year later from a surface of a centrosymmetric material [2]. SHG has been employed in laser devices for frequency conversion, high-resolution microscopy for biological and medical sciences, and as a spectroscopic probe in a wide variety of fields, including for measuring the orientation of crystal grains, sensing the static electric field at the surface of materials and measuring asymmetry and crystal defects [3–5]. One of the reasons SHG is an extremely powerful spectroscopic tool is that, for centrosymmetric materials, it originates from surfaces or defects with near-vanishing contributions from the bulk of the sample [6]. This is because second-order nonlinear processes are allowed only in non-centrosymmetric materials, while at the interface of any two materials or at the location of a defect this inversion symmetry is naturally broken [7]. As a result, surface SHG microscopy is especially useful in unravelling surface sensitive effects, including in photovoltaics and probing electrical double layers effects in catalysis, where the role of surfaces is vital for nearly all aspects of device function [8,9].

As the amplitude of SHG is proportional to the square of the local (optical) electric field, it can be augmented using dielectric or plasmonic resonances that give rise to substantial local field enhancement [10]. In particular, orders of magnitude SHG enhancements have been reported in individual nanostructures and in metasurfaces [11–20]. This in turn enables far easier observation of SHG from samples with weak nonlinear response. For instance, in nanoparticles, the angular dependence of SHG has been exploited to probe their symmetry, enabling a diagnostic tool for verifying whether a nanoparticle exhibits the expected shape and size [21]. However, while localized surface plasmon resonance (LSPR) enhanced SHG has now been extensively studied, enhancing SHG via delocalized, propagating surface plasmons has remained largely unexplored [22].

Two key issues prevent SHG, and especially SHG microscopy, being more widely applied as an interfacial probe. Firstly, SHG is a weak signal that, except with specialised set-ups optimized for a particular task (e.g. examining surface reconstruction) [8], can only be detected from a region close to the (sub-micron) excitation spot on a sample surface at any time, limiting its application for understanding wide sample regions. Secondly, when light is incident on a multilayer structure (for example a solar cell or battery) every interface and non-centrosymmetric (and possible other weak bulk effects [23]) can contribute to the SHG, making it challenging to disentangle competing contributions.

In this work we present delocalized SHG achieved through propagating surface plasmon polaritons (SPPs). Plasmon-enhanced SHG is observed as far as 35 μm away from the excitation spot and generated across a monocrystalline gold surface. This effect is facilitated by the interaction of counter-propagating SPPs with no or minimal concurrent emission of the light at the fundamental frequency. This yields SHG across a large monocrystalline area quasi-simultaneously (< ps) using solely point source illumination. In contrast to prior works on plasmonic nanoantennas or waveguides [24–28], which exhibited size-dependent response in terms of amplitude and spatial emission, we demonstrate a strong signal from unpatterned gold surfaces with nearly-uniform spatial intensity, regardless of the spatial separation between the gratings. This renders our method more suitable for generic dielectric-metal interfaces. Uniquely, we also demonstrate highly directional emission perpendicular to the gold surface, in stark contrast with SHG light scattered from plasmonic nanostructures over a wide range of angles [29]. The high signal strength allowed us to observe SHG light using a low-cost CMOS camera with 1 s exposure times and an industrial-grade pulsed laser with ~ 1 nJ per pulse (40 MHz repetition rate, <200 fs pulse width). We find that on a monocrystalline gold surface the SPP-induced SHG signal follows the same sensitivity to the local crystal structure as locally excited SHG. Our results prove that SPP-induced SHG can be employed as a surface probe of buried interfaces, resolving simultaneously the challenge of weak and localized SHG signals and thus enabling measurements in fields including material and surface research for energy, catalysis and others.

**Main**

Surface plasmon polaritons are optical excitation that travel along an interface between a materials with a negative real part of relative permittivity (typically a metal) and a dielectric [30]. SPPs can propagate significant distances (~ 200 µm $\frac{1}{e}$ amplitude decay length on a gold-air interface at 1030 nm, the wavelength relevant for this study) on flat metal surfaces provided the intrinsic Ohmic losses are sufficiently low [31,32]. SPPs, being evanescent surface waves, do not directly couple to the optical beams incident from the surroundings due to momentum mismatch, so their excitation is typically achieved via Kretschmann-type configuration or through employing a grating to provide additional momentum to the incident beam [30]. We employed the latter approach and used focused ion beam (FIB) milling to fabricate gratings for SPP excitation on a large {111}-surface of a thick (> 1 µm) monocrystalline gold flake (see Methods for synthesis and fabrication details [33]).

A schematic of the initially studied system is presented in Figure 1a: an incident pulsed laser beam is focused on an in-coupling grating, exciting an SPP that travels on the gold surface and is then out-coupled from an identical grating at 30 µm distance. The geometrical parameters of the grating were optimized for the operational wavelength of the laser – 1030nm, with details described in the Methods section, and optical image of the sample is shown in Figure 1b. As shown in Figure 1c, the gratings function as expected: illumination of the in-coupling grating (bright spot on the left-hand side) results in a visible signal at the out-coupling grating (two stripes on the right-hand side). We also carried out absorption measurements which suggest up to 20 % of the incident power couples to a surface-plasmon-polariton, as discussed in Supplementary Information Note 1.

We subsequently measured SHG from the same sample, shown in Figure 1d on a logarithmic scale. We confirmed the observed signal was SHG through both its spectrum and quadratic scaling with incident fundamental laser beam power (Supplementary Information Note 2). As expected, strong SHG is observed at a location coinciding with the excitation laser spot, due to the strong nearfields in the vicinity of the abrupt surface discontinuities at the grating. A weaker SHG signal is also clearly seen at the outcoupling grating, i.e. the signal is delocalized with respect to the excitation position, with total counts 0.5 % of the intensity of the SHG at the excitation spot, which can be clearly seen in a line profile presented in Figure 1e. Importantly, this plot also reveals that no SHG signal was detected from the planar gold surface between the gratings: overlaying the white light reflection clearly demonstrates that the SHG originates mostly from the ridges of the two gratings, in qualitative accordance with the previous theoretical considerations [34]. To the best of our knowledge, we report the first experimental demonstration of SPP-induced SHG with delocalization over > 10 μm length scales. To confirm our results we repeated this SHG measurements on three different nonlinear microscope setups and saw comparable results in all cases (see Methods). We also explored different grating spacings and saw comparable results for all grating spacings explored (10 – 35 μm).

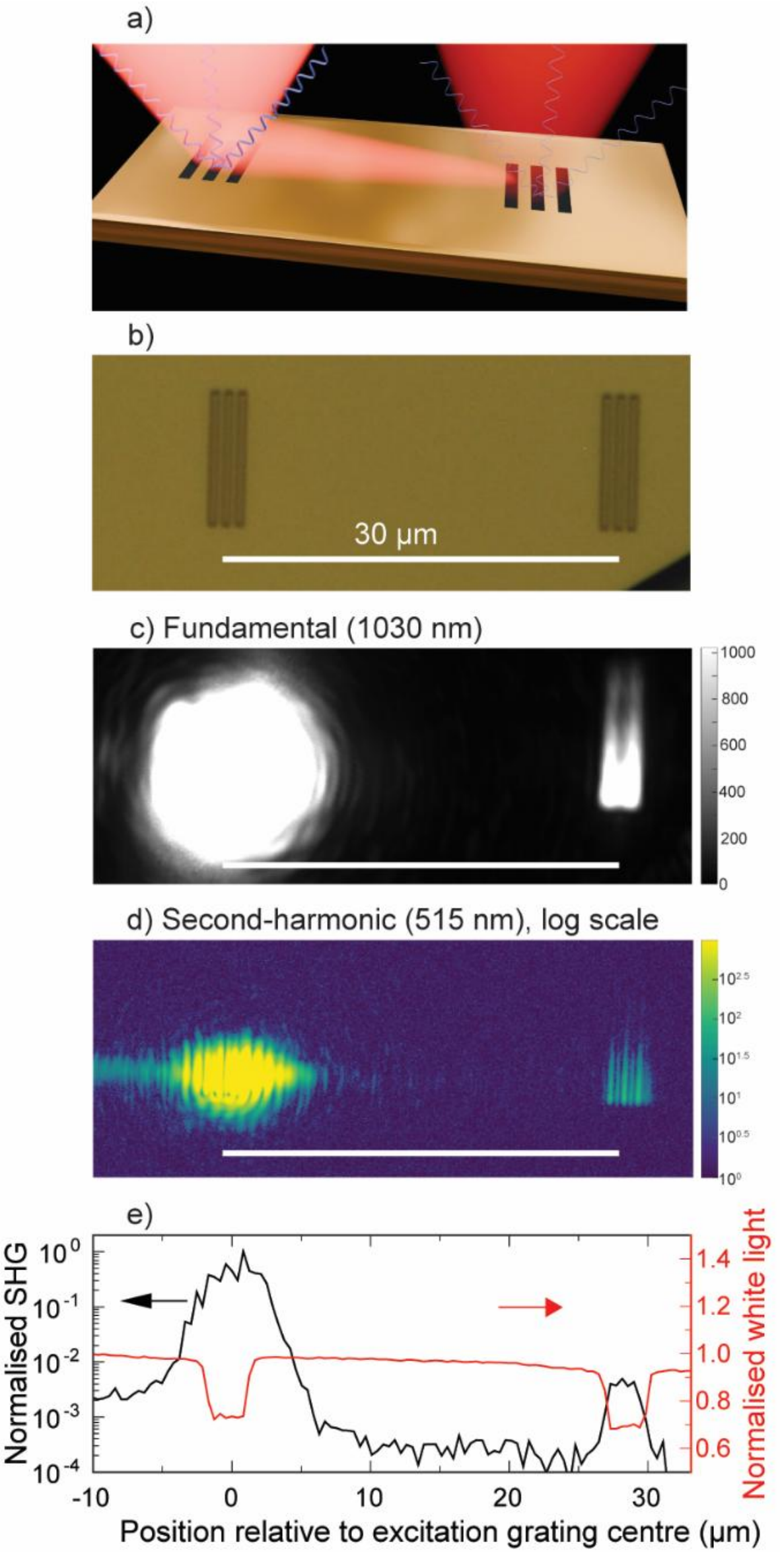

*Figure 1. a) Schematic illustration of the system: fundamental beam incident on a grating excites a surface plasmon polariton, which then out-couples through a second grating, marked in red. Fundamental marked in red and SHG in blue arrows. b) White light reflection image of the gratings fabricated on a monocrystalline gold flake. c)/d) Fundamental (1030 nm)/Second-harmonic (515 nm, logarithmic scale) images obtained upon illumination of the left grating with a focused laser beam. e) Second-harmonic and white light intensity (normalised to maximum) line profiles overlaid.*

We now explore producing delocalized SHG by counterpropagating SPPs directly on a flat surface, i.e. without exploiting an outcoupling grating. The principle is illustrated in Figure 2a: following SPP excitation at the in-coupling grating (1, on the left-hand side) a second grating (2, right-hand side), functioning as a Bragg reflector for the SPP, is now used to redirect the SPP back towards the excitation spot. We confirmed this second grating acts as a reflector using finite element method modelling, see Supplementary Information Note 3, with our simulations indicating approximately 33 % of the SPP's power is reflected at 1030 nm. This creates two counter-propagating SPPs waves and allows us to study effects arising due to their interference. An image of the structure fabricated on a monocrystalline gold surface is shown in Figure 2b. Figure 2c shows the intense reflection of the fundamental light from the excitation spot and only weak scattering from the left edge of the reflection grating. The SHG signal from this structure is shown in Figure 2d on a logarithmic scale. Excitingly, in addition to the SHG signal from the two gratings, similar to the described-above measurements, we detect SHG emission from the unpatterned surface between them. At the same time, we do not observe any significant fundamental beam emission from the region between the two gratings, apart from some minor speckles due to the scattering of the excitation laser beam. This is further demonstrated in the line profile in Figure 2e, where the fundamental and SHG signals are overlaid, both normalised to their strength at the outcoupling grating. While the fundamental signal drops to less than 5 % of this maximum just a few microns away from the grating, the SHG signal remains larger than 30 % of the maximum across the entire unpatterned region. This implies that the interference of the two counter-propagating SPPs, which are confined at the interface at the fundamental frequency, results in SHG that is observable in the far-field (see discussion of the physical mechanism below). We observe similar results for reflection gratings spaced from 10 µm to 35 µm from the in-coupling grating, as is presented in Supplementary Information Note 4. Thus, we demonstrate not only spatial decoupling of the SHG and fundamental signals on the surface, but also conversion of the intrinsically evanescent wave to a free-space propagating wave, which is not attainable with a single SPP, as demonstrated above. As we show next, this opens unprecedented opportunities for the study of buried interfaces.

While SHG is a powerful probe for studying single interfaces, in multilayer structures it can be challenging to assess which interfaces contribute to the signal, as noted above. We now demonstrate how delocalized, SPP-induced SHG can access a specific buried interface without use of non-centrosymmetric materials [35]. To this end, following the schematic in Figure 2f, we again fabricated in-coupling and reflective gratings on the monocrystalline gold surface, but this time we also sputtered a ~2 µm-thick $SiO_2$ layer on top of gold, creating a buried gold/glass interface. A white-light image of the resulting structure is presented in Figure 2g. As observed in the prior experiment, the signal at the fundamental wavelength is still strongest at the in-coupling grating (Figure 2h). Remarkably, the SHG signal is clearly visible over the entire region of interest even on a linear scale and the strongest SHG signal is now observed at the reflection grating (Figure 2i), located 20 µm away. This is further emphasised in the line profiles in Figure 2j. We also note that all SHG signals were significantly enhanced relative to those in air (for the region between the gratings, approximately by a factor of 5 in intensity), to the point that we could record the image in Figure 2i using an inexpensive colour CMOS microscopy camera with 1 s exposure time. SHG has been previously shown to be enhanced by dielectric layers on metals due to more favourable Fresnel reflection/transmission coefficients at interfaces [36]. Here we suggest further improvements are due to increased SPP vertical confinement, and hence the electric field amplitude, and refraction at the $SiO_2$ interface ensuring more of the beam is incident close to perpendicular to the in-coupling grating, though note that changes to the effective nonlinear susceptibility components are likely also important [37,38]. We believe this to be a novel report of SHG signal increasing in strength away from the excitation spot, and, importantly, the first report of spatially delocalized, SPP-induced surface-SHG from a wide (tens of micrometers) planar buried interface. While ensuring that the SHG signal originates solely from the interface of interest, this also enables the possibility of non-linear imaging without the need for a broad laser illumination area.

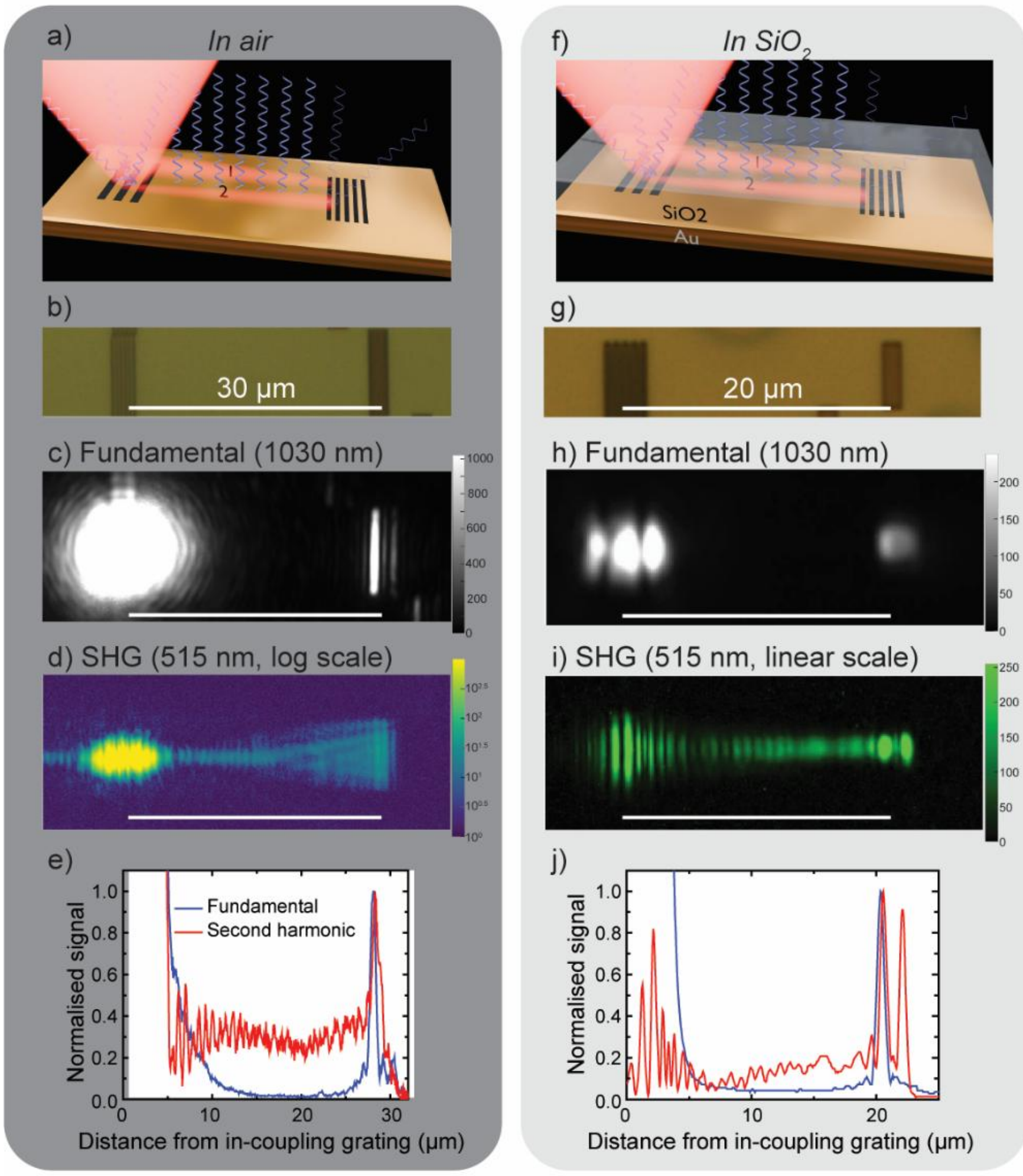

*Figure 2. a) Schematic of the system for excitation of counter-propagating SPPs and observation of SHG from a plain gold surface: fundamental beam incident on the grating excites an SPP that propagates from left to right (1) and gets back-reflected by the second grating (2), thereby creating two counter-propagating SPPs. Fundamental marked in red and SHG in blue arrows. b) White light reflection image of the fabricated sample with excitation and reflector gratings. c)/d) Fundamental/second-harmonic (linear/logarithmic scale) images obtained upon excitation at the fundamental grating. e) Fundamental and second-harmonic line profiles overlaid. f)-j) are the equivalent of a)-e) for the case when a layer of $SiO_2$ is deposited on the gold surface, noting that i) can now be plotted on a linear rather than logarithmic scale and that the grating geometrical parameters have been optimized for a gold/$SiO_2$ interface.*

We now study the physical mechanism behind the SPP-induced SHG from unpatterned surfaces. We recorded the back-focal-plane (i.e. angle resolved) SHG signal when i. exciting and detecting at a bare gold surface, ii. exciting and detecting at an in-coupling grating, and iii. exciting an in-coupling grating and detecting the region between an in-coupling and detection grating, as is presented in Figure 3a. For measurements ii. and iii. we placed an f-stop in our collection path to spatially filter the SHG signal, restricting SHG collection from the relevant surface region, as is presented in Supplementary Information Note 5 for the region between gratings. The maximum acceptance angle of our objective (NA=0.7) is approximately 44°. We present line profiles of back-focal-plane results for the three cases in Figure 3b. For case i. and ii. we observe contributions to the SHG signal at a wide range of angles. Notably, for case ii. emission is at 27±1°, which compares well with the predicted value for a grating of 30.6°. Conversely, for case iii. the SHG signal is concentrated around the centre of the back-focal-plane, indicating that SPP-induced SHG is preferentially emitted perpendicular to the gold surface (see Figure S5c for a 2-dimensional image of the back-focal-plane in this case). This is suggestive of k-space restrictions on how SPP-induced SHG is formed.

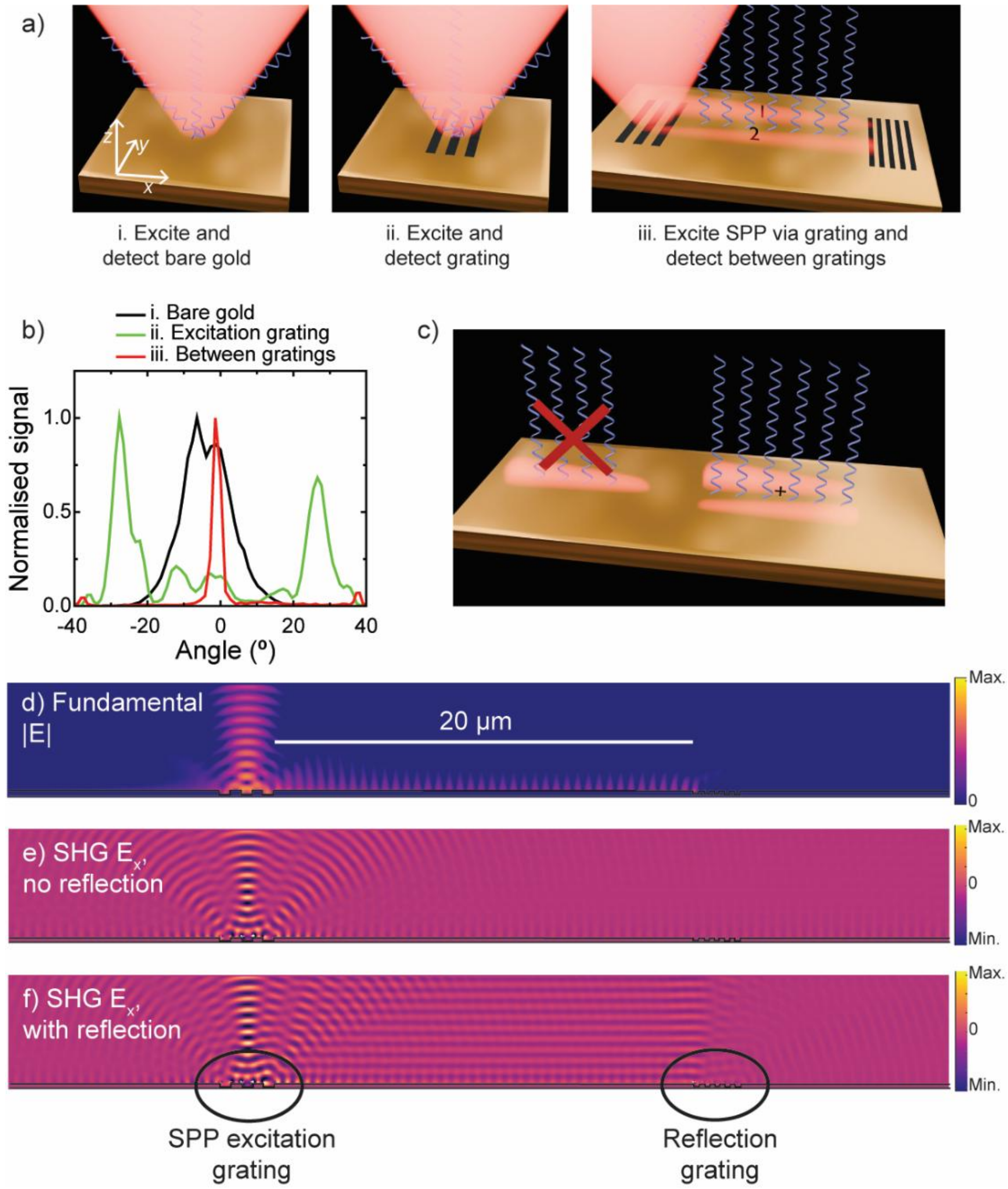

*Figure 3. a) Schematic demonstrating three measurement configurations: i. excitation and detection on a bare gold surface, ii. excitation and detection on a grating, and iii. excitation of an SPP via a grating and detection in the region between gratings. b) Line profiles of angle resolved second-harmonic*

*signals for cases i., ii. and iii. (with each signal normalised to its maximum). Measured at a gold/air interface. c) Schematic of processes – a single surface plasmon polariton cannot couple to free space via second-harmonic generation, while two counter-propagating surface plasmon polaritons can produce second-harmonic signal perpendicular to the surface. d) Simulated distribution of the logarithm of the normalised fundamental electric field, for a Gaussian beam excitation centred on an in-coupling grating. e) and f) Simulated distribution of the second-harmonic in-plane electric field parallel to the gold surface produced by the excitation shown in d), without and with the reflection grating (on the same scale). x coordinate direction marked in a).*

We developed an analytical model of SHG from SPPs, which we present fully in the Supplementary Information Note 6. Briefly, we find that for a single SPP any SHG waves that are produced remain evanescent, i.e. they do not outcouple from the surface to the far field. We note that similarly to the linear case, the nonlinear emission can be also out-coupled to the far-field using a prism (Kretschmann configuration) [17]. However, in the case of two counter-propagating SPPs, one has an electric field proportional to $E_{\mathrm{SPP},1} \propto e^{i(\beta x - \omega t)}$ (considering wave propagation, which is the same for all field components) while the second is proportional to $E_{\mathrm{SPP},2} \propto e^{i(-\beta x - \omega t)}$. Here $\beta$ is the in-plane SPP wavevector, $x$ the coordinate along the direction of propagation of the SPP and $t$ time. As SHG is proportional to the electric field squared (i.e. $\left(E_{\mathrm{SPP},1} + E_{\mathrm{SPP},2}\right)^2$), for two counter-propagating SPPs this expression contains a term that is constant in $x$ (the resulting terms are often re-written in the form of a standing wave, but the frequency components remain a constant and oscillating term). An optical electric field at frequency $2\omega$ which is constant in amplitude as position is varied across the sample surface is, by definition, a plane wave perpendicular to the sample surface. A schematic of this concept is presented in Figure 3c and we note that this process was proposed by Fukui et al. [39] (though using a different modelling approach) and later observed in macroscopic experiments [40,41]. Our model reveals that the nonlinear polarisation component responsible for the SHG must be an in-plane polarisation (in this situation $\vec{P}_{\mathrm{x}}$, see discussion below), while the out-of-plane polarisation cannot contribute to this phenomenon. Finally, we are not aware of other SHG-based optical processes that so selectively radiate into only one angle.

To further confirm our analytical model, we carried out finite element method simulations for the grating structure presented in Figure 2a with and without the reflection grating present (see details in the Methods section). The resulting fundamental for the case with the reflection grating is shown in Figure 3d, demonstrating an incident Gaussian focused beam (within the paraxial approximation) exciting SPPs and radiation only coupling to the far-field originating from gratings. The modelled SHG, for the case without and with a reflection grating, are presented in Figure 3e and f respectively. Specifically, we plot the electric field amplitude parallel to the surface ($E_x$) as this highlights waves propagating away from the gold. For the case without a reflection grating, SHG only couples to the far-field at the gratings, comparable to the fundamental beam. Conversely, when a reflection grating is present (Figure 3f), SHG is produced both at the gratings (at a wide range of angles, in agreement with our measurements in Figure 3b) and, importantly, across the entire unpatterned monocrystalline surface, where the beam is emitted only in the surface-normal direction. We further confirmed our model by simulating a single SPP travelling across a monocrystalline surface and observed no coupling to far-field SHG, while two counter-propagating SPPs do couple to far-field SHG (see Supplementary Information Note 6). Finally, we note that in our experimental results the SHG image exhibits some fringes (note the oscillatory features in Figures 2e and i), which we do not expect from simulations. These oscillations make up to ~ 20 % of the total beam intensity. We present a thorough investigation of these features in Supplementary Information Note 7, demonstrating that they originate from interference of the SPP-induced SHG and SHG originating from the gratings, which has a small component far from the gratings due to the finite numerical aperture of the microscope.

Finally, thanks to the use of monocrystalline gold surfaces ($\{111\}$ orientation), we can investigate which nonlinear susceptibility components produce the SHG signal (noting that this phenomenon could result from many different crystal surfaces and susceptibility components). Importantly, in comparison with an isotropic surface (e.g. evaporated or sputtered polycrystalline films), a $\{111\}$-type crystalline gold surface features additional anisotropic elements of the second-order susceptibility tensor $\overleftrightarrow{\chi}^{(2)}$. This is due to isotropic surfaces having $c_\infty$ symmetry, whereas $\{111\}$-type surface belongs to $c_{3v}$ symmetry group [37,42]. Hence, in the case of an isotropic surface, only interaction with a surface-normal polarized optical field would result in SHG, since the inversion symmetry is broken only across the interface. In turn, $\{111\}$-type surface additionally has broken inversion symmetry in-plane, and thus possess additional $\overleftrightarrow{\chi}^{(2)}$tensor elements that also interact with surface-parallel polarized optical field.

All gratings presented thus far in our text have been parallel to crystal edges (i.e. along a $\langle\bar{1}10\rangle$-type axis), as shown in Figure 4a (orange circle, gratings are parallel to the edge of the flake). We fabricated a second set of gratings perpendicular to this crystal axis, which corresponds to $\langle 11\bar{2}\rangle$-type axis (Figure 4a, blue circle). In both cases, we studied the polarization of the fundamental and SHG signals by introducing a polariser on the collection path, rotating it and recording the change in the signal intensity. As SPP excitation requires the fundamental beam to be polarised perpendicular to the grating grooves, for both the parallel and perpendicular grating orientations we observe a typical cosine-squared response for the fundamental beams (Figures 4b and c, black squares). We then studied the SHG, again using the f-stop to spatially select the signal only from the unpatterned region of the monocrystalline surface between the gratings. For the grating parallel to the crystal edge, the SHG polarisation follows the same polarisation dependence as the fundamental beam (Figure 4b, red circles). However, for the gratings perpendicular to the crystal edge the SHG beam is cross-polarized with respect to the fundamental beam (Figure 4c, red circles). This has previously been observed in local SHG measurements on {111} monocrystalline gold surfaces and demonstrates that the SHG signal is produced by a nonlinear susceptibility component sensitive to the crystal facet [36,43]. Based on these results and our modelling (see Supplementary Information Note 6), we identify that our signals are due to $\chi^{(2)}_{xxx}$ and $\chi^{(2)}_{\mathrm{xyy}}$ for SPPs launched by gratings parallel to and perpendicular to a crystal edges respectively (where the angle between $x$ and $y$ is 90°). Therefore, we have shown that in-plane components of the second-order susceptibility tensor dominate the SHG response in this case, which is also consistent with the previous reports. Out-of-plane components, predominantly $\chi^{(2)}_{zzz}$ (as well as other symmetry-allowed tensor elements containing $z$ subscript) would be of primary significance for polycrystalline samples, which may also give rise to measurable SHG by counterpropagating SPPs [40].

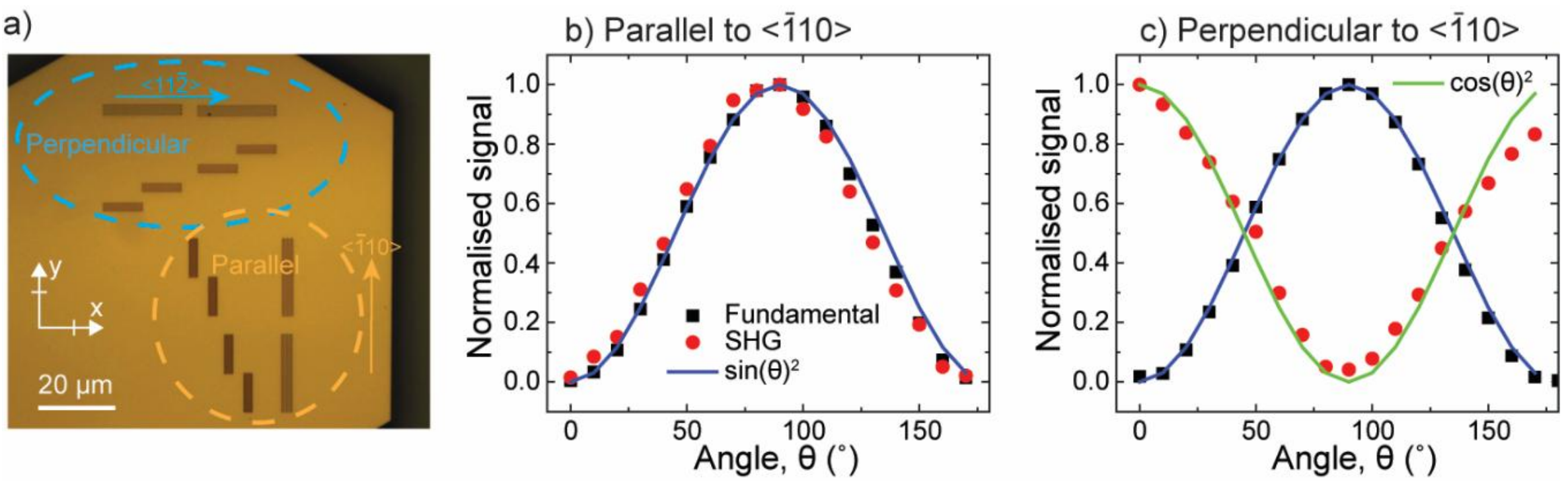

*Figure 4. a) White light reflection image of the fabricated sample with in-coupling and reflection gratings aligned parallel to the crystal edge (i.e. along ⟨$\bar{1}10$⟩-type crystal axis), and perpendicular to the crystal edge (i.e. along⟨$11\bar{2}$⟩-type axis), as marked on figure. b)/c) Outgoing wave polarisation dependence of the fundamental (full sample) and second-harmonic (measured in the region between gratings) signals for gratings aligned parallel to/perpendicular to gold crystal planes. sin and cosine squared functions are plotted for comparison. Legend in b) applies to b) and c).*

**Conclusion**

In summary, we have demonstrated delocalized SHG at up to 35 μm from the excitation beam by coupling the fundamental beam to SPPs. More excitingly, we reveal that SHG can be generated from a flat monocrystalline surface, with minimal concurrent far-field fundamental emission, through the interaction of two counter-propagating SPPs. We believe this to be the first time such signal has been observed at the microscale. We show that emitted SHG from the monocrystalline surface is highly directional, propagating perpendicular to the interface only, and that the polarisation of this signal is sensitive to the local crystal structure. Furthermore, we demonstrate that this signal can arise from a single buried interface within a multilayer structure and can be recorded with an inexpensive CMOS camera, industrial-grade laser and reasonable (~1 s) exposure times. We believe that our results open new possibilities in fundamental and applied research, demonstrating possibilities for delocalized enhancement of SHG and other nonlinear effects across buried interfaces, and paving the way to wide variety of applications including in multilayer structures for energy applications, surface science and biological imaging.

## Methods

*Gold flake synthesis:* Chemical synthesis procedure followed the same methodology as described elsewhere [44,45]. Briefly, Au flakes are grown directly on a Si substrate by reducing a gold salt precursor ($HAuCl_4$) dissolved in organic solvent (toluene) via thermolysis. Due to random nature of the synthesis process, a large variety of gold crystals are created in every synthesis batch and only the large, clean and undamaged samples are selected for subsequent patterning using FIB.

*Designing in-coupling gratings:* Gratings were designed following the approach outlined in [30]. Specifically, the grating period was chosen to have the same spatial frequency as a surface plasmon polariton travelling along the surface of a gold flake. Thus, when a beam is incident perpendicular to the structure it is diffracted directly into an SPP channel. Based on previous studies [46] we chose the depth of the etched part of the gratings to be approximately 70 nm, and half of the region between each period was etched using focused ion-beam milling. We used a grating period of 1010 nm and 469 nm for SPP excitation reflection gratings at a gold-air interface, and 696 nm and 280 nm for a gold-$SiO_2$ interface.

*Focused Ion Beam Milling:* Following the chemical synthesis, the designed grating patterns were FIB-milled on the selected gold flake samples. The fabrication was performed using a Zeiss Crossbeam 540 dual beam instrument, at 30 kV acceleration voltage and 10 pA ion beam current. The desired milling depth (approximately 70 nm) was achieved by optimizing exposure dose (we performed pre-characterization of the milling rate at the given beam parameters, and obtaining 0.02 nC/µm$^2$ as the optimal value). Scanning electron microscope images of the patterned flakes are provided in Supplementary Information Note 8. We additionally note that we initially tried milling gratings on thin gold flakes on glass substrates (using a thin layer of graphene to prevent charging), but this resulted in significantly weaker coupling to surface plasmon polaritons.

*$SiO_2$ deposition:* Following the FIB fabrication, an approximately 2 µm-thick layer of SiO2 was RF-sputtered on the gold flake sample using Pfeiffer SPIDER 600 deposition system (35 minutes at 1 kW power, with argon and oxygen inflow into the chamber during the deposition).

*Second-harmonic generation microscopy:* All measurements presented were carried out on a NanoMicroSpec-Transmission™ (NT&C) microscope, further adapted to enable second-harmonic generation microscopy. Specifically, a <200 fs pulsed -wave laser (1030 nm, 40 MHz, pulse energy ~ 2.5 nJ, NKT Photonics Origami 10-40) was coupled into a Nikon Eclipse Ti2 inverted microscope. Here the objective was a 60x Nikon S Plan Fluor (NA=0.7). The laser beam was passed through a dichroic shortpass mirror (Edmund Optics, 750 nm cut-off wavelength) prior to impacting on the sample. When imaging the fundamental beam neutral density filters (Thorlabs) were placed in the collection path to reduce the signal strength. For second-harmonic generation images two 850 nm shortpass filters (Thorlabs FESH0850) and one bandpass filter (Thorlabs FBH515-10) were placed on the collection path to remove any fundamental signal, while for spectroscopic measurements only the shortpass filters were used. A 4-f collection system was constructed before our spectrometer and camera to allow for spatial and angular filtering. For images of fundamental and SHG beams we employed a CMOS camera (Thorlabs CS165MU/M) after our 4-f collection system, while white-light reflection images used the in-built (colour) NT&C camera. The signal was coupled to a Princeton Instruments Spectra Pro HRS-500 spectrometer and recorded on a PIXIS 256 camera for spectroscopic measurements. Finally, we note that our laser was slightly diverging so we used an f=400 mm achromatic lens to collimate the beam (Thorlabs AC254-400-B). The laser spot size was recorded using a camera and fitted by a two-dimensional Gaussian. Spot size here is as the intensity that falls to $\frac{1}{e^2}$ of the maximum intensity. In all measurements we used a Gaussian shaped beam with a spot radius of 3 ± 0.2 μm that filled the central portion of the back-focal-plane (as gratings couple most effectively to light incident perpendicular to the sample), and unless otherwise stated the beam power on the sample was 40 ± 5 mW. A schematic of the setup is shown in Supplementary Information Note 9.

A portion of experiments were repeated on two different nonlinear microscopy setups, equipped with a Ti:Saph and Mai-Tai lasers both operating at 820 nm and 1030 nm respectively. The obtained results were qualitatively similar, although with different relative intensities and different fringe periodicities were observed in the counter-propagating SPP experiments.

*Laser Power Determination:* was recorded using a Thorlabs S170C or S130C power meter (using the microscope power meter for any measurements at the sample position).

*Back-focal-plane imaging:* Two lenses were placed between the image plane of the microscope and the spectrometer/camera to enable back-focal-plane measurements, following methods described by Kurvits et al. [47].

*COMSOL modelling:* All modelling presented in the Main text and Supplementary Information was carried out using the finite element modelling method in the wave optics package. McPeak's experimental dielectric constant of gold was used in all simulations [31]. All simulations were two dimensional and had a perfectly matched layer surrounding all sides of the simulation cell. Nonlinear polarisation was defined in the same way as this tutorial [48], but the region of polarisation was chosen to be 20 nm thick directly above the gold surface, representing polarisation at a surface. We confirmed that this approach produced surface-based second-harmonic generation results correctly prior to applying it to our application: we showed that a Gaussian beam focused on the surface produced the expected angular distribution of SHG for different nonlinear susceptibility terms. All COMSOL results presented are for a fundamental beam at 1030 nm and the grating spacing is identical to that used in our measurements. We used $x$ polarised second-harmonic nonlinear polarisation in all simulations presented (where $x$ is the direction parallel to the gold surface), with the nonlinear polarisation $\vec{P}_x^{nls} \propto E_x E_x$ in all simulations except Figure 3e and f, where we present results for $\vec{P}_x^{nls} \propto E_z E_z$ (with $z$ perpendicular to the gold surface). While we see comparable results for both polarisation terms in this simulation (and note that we demonstrate above that the $E_x E_x$ is the key component observed in experiments), the $E_x E_x$ results are more strongly dominated by SHG emission from the gratings, preventing easy observation of the emission between the gratings in simulations. Therefore we present the result that gives the clearer image (noting that our simulations are likely to be inaccurate in magnitude near gratings due to fabrication limitations). We further confirmed that the region between the gratings gives SHG emitted perpendicular to the surface for the $\vec{P}_x^{nls} \propto E_x E_x$ option by carrying out a second simulation in which we excluded the gratings from the nonlinear polarisation layer above the gold, and our results are fully consistent with results in this text. Finally, nonlinear simulations were verified using a method described by Reddy et al. [49], yielding qualitatively similar results. However, this method exhibits worse numerical convergence, as calculation of the nonlinear polarization involves tangential derivative of the electric field, which has abrupt discontinuity at the sharp features of the grating structure.

**Acknowledgements**

ARB acknowledges funding from SNSF Swiss Postdoctoral Fellowship TMPFP2_217040. ARB and LMH acknowledge funding from the Engineering and Physical Sciences Research Council (EPSRC) UK. ARB and GT acknowledges the SNSF Eccellenza Grant PCEGP2-194181.

**Conflicts of Interest**

There are no conflicts of interest.

# Supporting Information: Plasmon Induced Delocalized Second-Harmonic Generation Towards Buried-Interface Spectroscopy

**Alan R. Bowman[1,2], Sergejs Boroviks[3], Omer Can Karaman[1], Laura M. Herz[2], Olivier J.F. Martin[3] and Giulia Tagliabue[1+]**

1. Laboratory of Nanoscience for Energy Technologies (LNET), STI, École Polytechnique Fédérale de Lausanne (EPFL), Lausanne 1015, Switzerland
2. Clarendon Laboratory, Department of Physics, University of Oxford, Parks Road, Oxford, OX1 3PU, UK
3. Nanophotonics and Metrology Laboratory (NAM), Swiss Federal Institute of Technology Lausanne (EPFL), 1015 Lausanne, Switzerland

+corresponding author: giulia.tagliabue@epfl.ch

*Supporting Information Note 1 – reflection measurements on gratings*

We carried out white light reflection measurements on the in-coupling gratings, referencing our signal to that of bare gold, for light incident with polarisation both parallel and perpendicular to the grating axis (following methods in our previous work) [1]. Results are presented in Figure S1. As expected, when the electric field polarisation is perpendicular to the grating there is significantly less reflection, due to the excitation or surface plasmon polaritons. More specifically, our measurements show a broadband reduction in reflection between ~700 nm and ~ 1100 nm, compared to the parallel polarisation case. We attribute this to our gratings being square, meaning they include many optical frequencies. We designed our gratings to excite surface plasmon polaritons at 1030 nm. At this wavelength, our measurements suggest up to 20 % of the incident beam couples to surface plasmon polaritons.

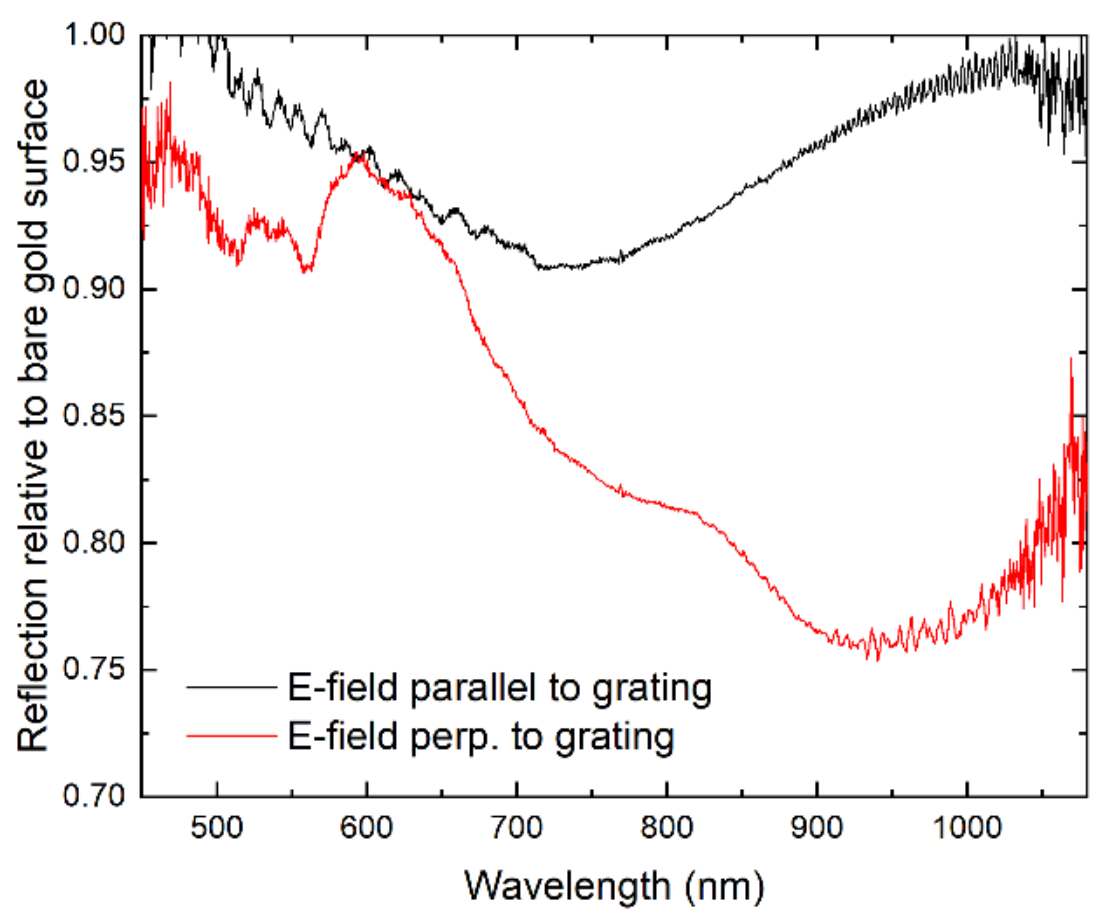

*Figure S1. Reflection of light incident perpendicular to the sample plane, relative to reflection from a bare gold surface, for the incident electric field being polarised parallel (black) and perpendicular (red) to the grating.*

*Supporting Information Note 2 – confirmation of second-harmonic generation*

We confirmed we are observing second-harmonic generation through spectral and power-dependant measurements, as presented in Figure S2. For all samples measured we recorded the spectrum of the detected signal after a 750 nm dichroic shortpass mirror and two 850 nm shortpass filters (Methods). In Figure S2a, we present these results both for regions surrounding the grating and for the signal between gratings (when employing a reflection grating). It can be seen both signals have a strong peak at the SHG wavelength (515 nm) and a weak and broad two-photon photoluminescence signal at longer wavelengths, consistent with previous reports [2]. In spatially resolved measurements we used a 515 nm bandpass filter to remove the contribution from two photon photoluminescence (Methods). We also recorded the power scaling of the SHG signal with laser power, and present our results in Figure S2b on a log-log scale. We fit with a straight line and obtain a coefficient of 1.95 i.e. implying counts are quadratic with laser power, confirming SHG.

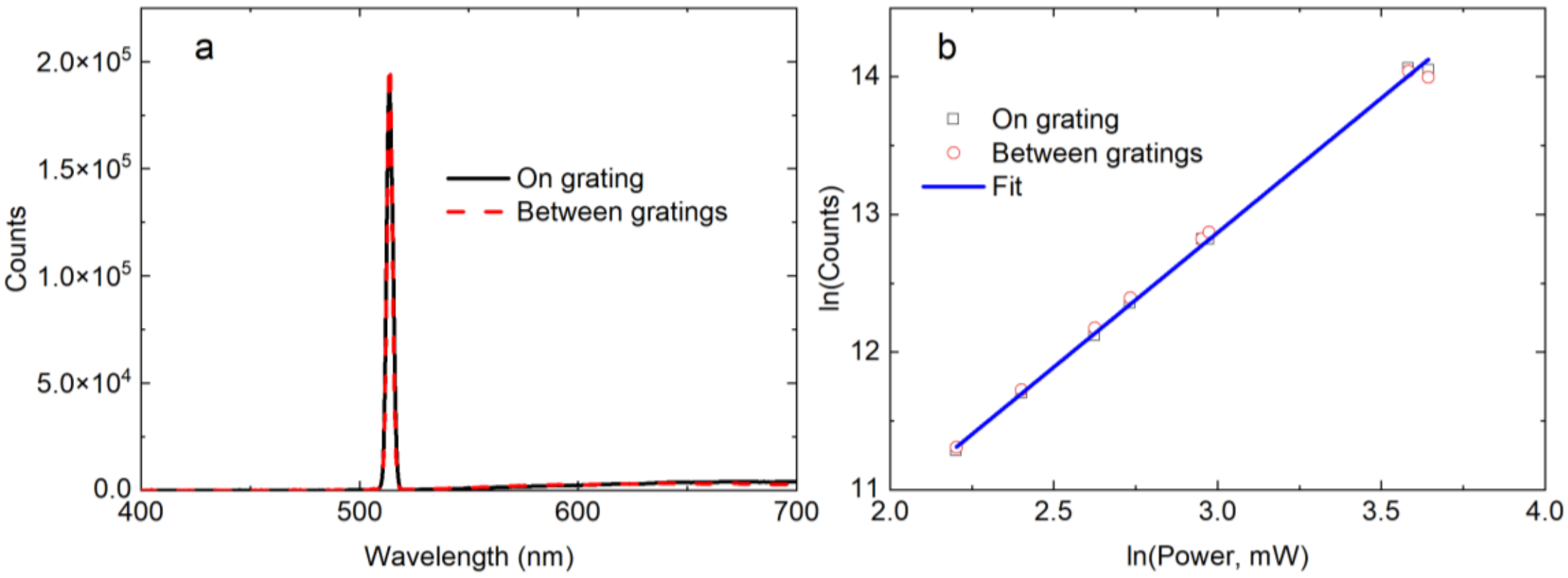

*Figure S2. a) Spectrally resolved SHG signal and b) log-log scaling of signal at 515 nm with laser power, for both a region surrounding an in-coupling grating, and the region between an in-coupling grating and a reflection grating. Data presented here is for a gold/$SiO_2$ interface.*

*Supporting Information Note 3 – finite element method modelling of reflector grating*

To confirm the role of our reflection gratings, we used finite element method modelling of a gold-air interface. We analytically defined the SPP travelling across the surface, and we present the electric field amplitude perpendicular to the surface in Figure S3a. We then recorded how this field changes as 'reflection gratings' are introduced into the simulation. By comparing Figure S3a (no grating) and b (with grating), the electric field amplitude prior to the gratings (noting SPPs are incident from the left side of the simulation) is seen to increase in the latter case. Conversely, in Figures S3c and d we present the time-averaged power flow of the electric field in the original direction of SPP propagation without and with the grating present. When the grating is introduced we observe a reduction in the power flow (a reduction of 30-40 %). This is explained by a significant portion of the SPP being reflected by the grating, reducing the time averaged power flow while increasing the field amplitude (i.e. a partial standing wave). By placing a port in our simulations at the left hand edge of the simulation and recording the time averaged power flow through it both with and without the grating, we were able to calculate that approximately 33 % of the power is reflected into a backwards propagating SPP by the grating.

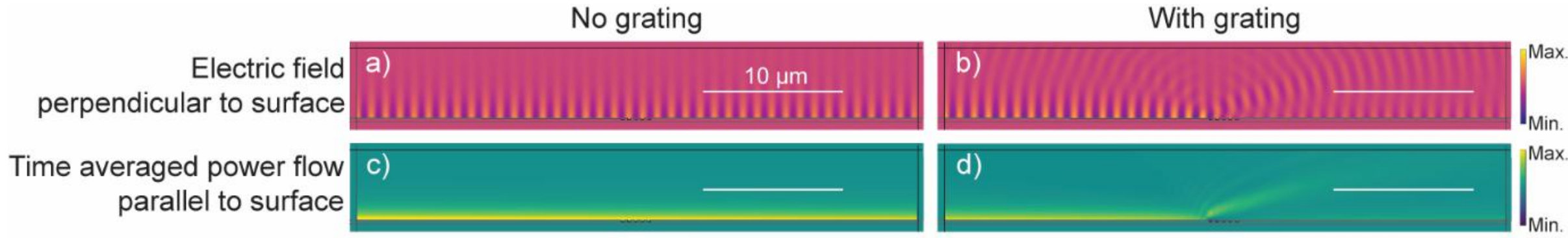

*Figure S3. COMSOL simulations of a surface plasmon polariton travelling along a gold/air interface from left to right. Specifically, the surface perpendicular component of the electric field amplitude (a, b) and the surface parallel time averaged power flow (c, d) for a gold/air interface are shown for the case without (left) and with (right) a reflection grating in the simulation. Outer boundaries in all simulations are perfectly matched layers.*

*Supporting Information Note 4 – distance dependence of SHG between gratings*

In Figure S4a we present a monocrystalline flake with an in-coupling grating between two reflection gratings, with the distance between the gratings varied across the sample (from 10 µm to 35 µm). In Figure S4b-g we present the observed SHG when exciting the in-coupling grating at different positions, to excite each structure in turn (with the letters b-g marked in Figure S4a to denote which grating is being measured). The exposure time is the same in all plots (5 s) and the length scale is the same on all plots. As can be seen, we observe qualitatively similar results in all cases.

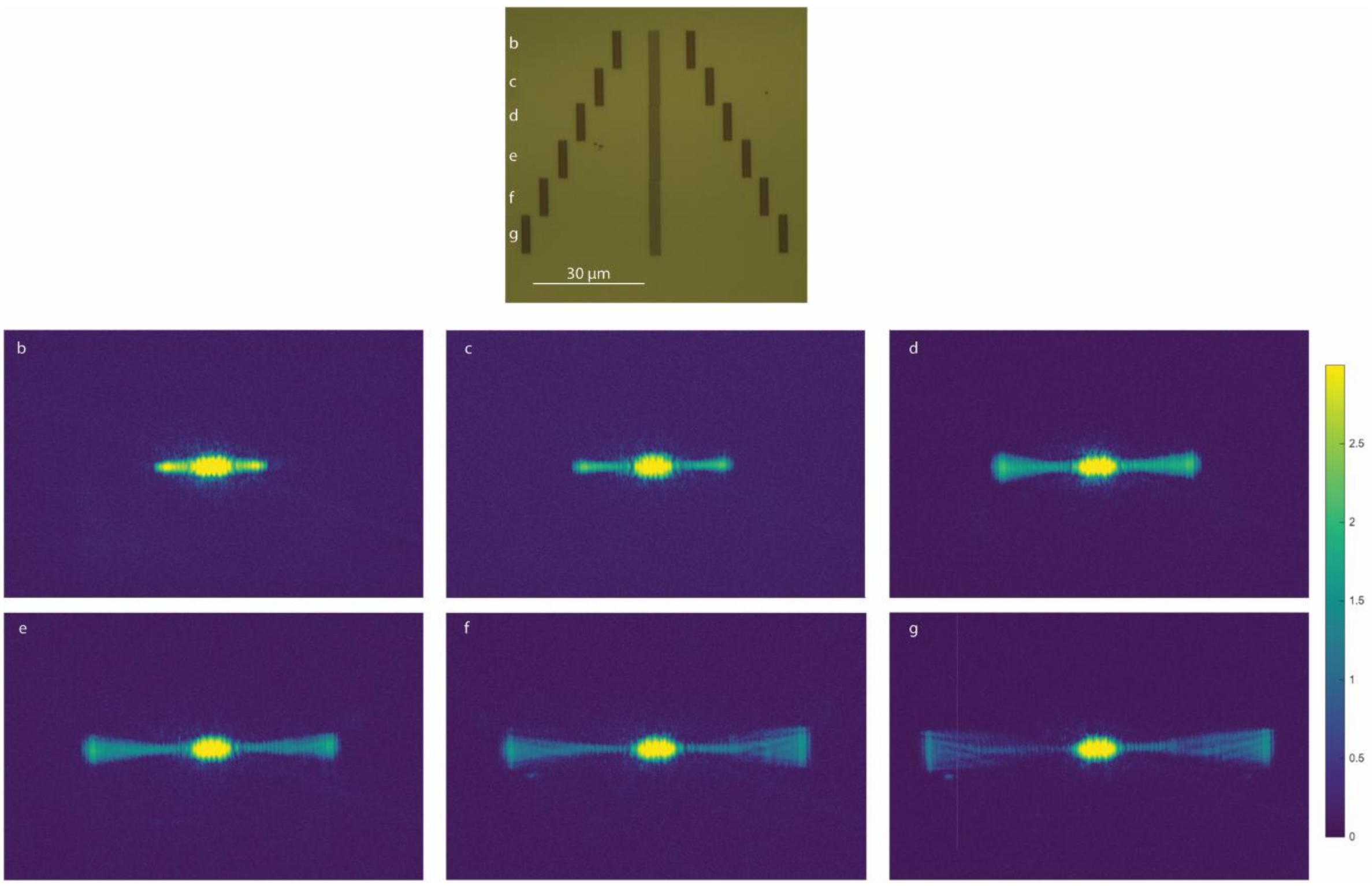

*Figure S4. a) White light reflection image of a flake with a series of reflection gratings at different distances from the in-coupling grating. Each grating structure is marked b)-g), and the SHG from each structure is presented in the subsequent parts of the image. Scale bar is the same on all plots.*

*Supporting Information Note 5 – back-focal-plane measurement details*

Figure S5 presents the approach taken for measuring spatially filtered back-focal-plane images in this manuscript, when considering emission between two grating structures. Specifically, for the SHG signal observed in Figure S5a we introduce an f-stop on the collection path to spatially signal the signal region from the flat surface, as presented in Figure S5b (blue dashes represent maximum spatial region measured). We then record the back-focal-plane signal from this spatially filtered region, as presented in Figure S5c. The orange dashed line represents the maximum extent of the back-focal-plane and the red dashed line shows how the line profile was taken to produce the results in Figure 3.

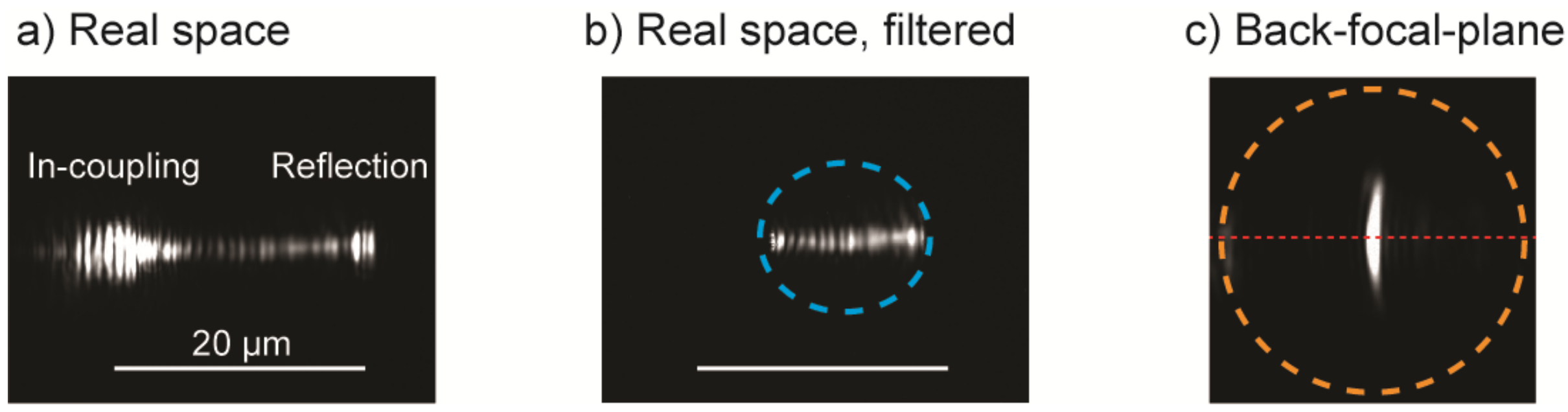

*Figure S5. a) Second-harmonic image when exciting a fundamental grating (left-hand side) and SPP is directed to a reflection grating (right hand side), similar to Figure 2i. b) Equivalent image when an f-stop is closed (dashed blue line), spatially filtering the light from the monocrystalline surface. c) Back-focal-plane of image b), with dashed orange line showing maximum extent of back-focal plane (~ 44°).*

*Supporting Information Note 6 – analytical and numerical modelling of SHG from counter-propagating SPPs*

*i) Analytical modelling*

We first introduce results for the waves emitted from a nonlinear polarization at a surface, derived by Heinz [3], before considering the specific case of a surface plasmon polariton generating this polarization. We consider the nonlinear polarization $\vec{P}^{nls}$ at a plane between two materials defined by $z = 0$ (see Figure 3 for axis definition), where we can write

$$\vec{P}^{nls} = \vec{P}_s^{nls}(x, y, t)\delta(z), \tag{1}$$

where

$$\vec{P}_s^{nls}(x, y, t) = \vec{P}_s e^{i(px-\Omega t)}. \tag{2}$$

Here $\Omega$ is the angular frequency of the oscillation and we have assumed uniformity in the $y$ direction. Heinz shows that from this polarization there will be two emitted waves, one travelling in each of the positive and negative $z$ directions. We consider a two-dimensional model (x-z plane) and we study the wave emitted in the positive $z$ direction. Following Heinz, the wave-vector of this wave is defined as

$$\vec{k}_1 = p\hat{\hat{x}} + q_1\hat{\hat{z}}, \tag{3}$$

where $q_1 = (\epsilon_1 K^2 - p^2)^{\frac{1}{2}}$, where $\epsilon_1$ is the permittivity in this $z > 0$ region, $K = \frac{\Omega}{c}$, where $c$ is the speed of light, and we have assumed $\mu = 1$. Furthermore, Heinz shows that the form of this emitted wave (in Gaussian units) is

$$\vec{E}_{nl} = \frac{2\pi i K^2}{q_1}\left(\vec{P}_s - \hat{\hat{k}}_1\left(\hat{\hat{k}}_1.\vec{P}_s\right)\right) \tag{4}$$

where $\hat{\hat{k}}_1 = \frac{\vec{k}_1}{|\vec{k}_1|}$.

*SPP travelling along the surface*

We now consider a single SPP travelling along the surface of a metal at $z = 0$ in the $x - z$ plane. Here we have [4]

$$\vec{E} = \begin{pmatrix} \frac{iBk_2}{\omega\epsilon_0} \\ 0 \\ -\frac{B\beta}{\omega\epsilon_0} \end{pmatrix} e^{i(\beta x-\omega t)}e^{-k_2 z} \tag{5}$$

above the metal ($z > 0$), with (for air above the metal) $\beta = k_0\sqrt{\frac{\epsilon_\mathrm{m}}{1+\epsilon_\mathrm{m}}}$, $k_0 = \frac{2\pi}{\lambda_0}$ i.e. the free-space wave-vector at wavelength $\lambda_0$, $\epsilon_\mathrm{m}$ the permittivity of the metal, $k_2 = (\beta^2 - k_0^2)^{\frac{1}{2}}$ (noting that $\beta$ is a complex number) and $B$ an arbitrary constant determining the strength of the SPP. Considering the nonlinear polarization results from the interaction of two electric field components (i.e. $E_iE_j$, second-harmonic generation) then we can state that for all nonlinear polarization terms

$$\vec{P}_i^{nls} \propto \sum_{j,k} \chi_{ijk}^{(2)} E_j E_k \propto e^{2i(\beta x - \omega t)}. \tag{6}$$

where $i, j, k$ refer to coordinate axes and $\chi_{ijk}^{(2)}$ is the second order nonlinear susceptibility for second-harmonic generation. Using equations 1-3 we see that $q_1 = \left(\frac{4\omega^2}{c^2} - \beta^2\right)^{\frac{1}{2}} = \frac{2\omega}{c}\left(1 - \frac{\epsilon_\mathrm{m}}{1+\epsilon_\mathrm{m}}\right)^{\frac{1}{2}} = \frac{2\omega}{c}\left(\frac{1}{1+\epsilon_\mathrm{m}}\right)^{\frac{1}{2}}$. Importantly, $\epsilon_\mathrm{m}$, the permittivity of the metal is always a complex number whose real part is less than -1 at optical frequencies, meaning that $q_1$ is a complex number. In other words SHG from a single SPP cannot couple to free-space as free-space wave-vectors are always real. Thus a single SPP can never produce far-field SHG, unless it can outcouple via other momentum-matching mechanism such as grating coupling (e.g. Kretschmann configuration).

*Counterpropagating SPPs on a surface*

We now consider a forward and backward travelling SPP, i.e.

$$\vec{E}_1 = \begin{pmatrix} \frac{iBk_2}{\omega\epsilon_0} \\ 0 \\ -\frac{B\beta}{\omega\epsilon_0} \end{pmatrix} e^{i(\beta x - \omega t)} e^{-k_2 z} \tag{7}$$

and

$$\vec{E}_2 = \begin{pmatrix} \frac{iBk_2}{\omega\epsilon_0} \\ 0 \\ -\frac{B\beta}{\omega\epsilon_0} \end{pmatrix} e^{i(-\beta x - \omega t)} e^{-k_2 z}. \tag{8}$$

If we consider the polarization resulting from interaction of the two electric fields (i.e. $E_{1,i}E_{2,j}$) we find that $\vec{P}_s^{nls} \propto e^{-2i\omega t}$ i.e. $\vec{P}_s^{nls}$ is no longer a function of $x$. This in turn gives $q_1 = \frac{2\omega}{c} = K$ and, more importantly, $\vec{k}_1 = q_1\hat{\hat{z}}$ i.e. the only free-space wave counter-propagating SPPs can couple to is one travelling perpendicular to the surface. This is what is observed in our experiments in the main text.

We now consider the polarization term in more detail. It was presented above that

$$\vec{E}_{nl} \propto \left(\vec{P}_s - \hat{\hat{k}}_1\left(\hat{\hat{k}}_1.\vec{P}_s\right)\right). \tag{9}$$

As we have just shown $\hat{\hat{k}}_1 = \begin{pmatrix}0\\0\\1\end{pmatrix}$, this in turn means that $\vec{E}_{nl}$ can have no $z$ component (this is clear as no plane wave can have an electric field component in the direction of travel). If we consider the coordinate system used in this text, where $x$ is the direction SPPs travel in when gratings are parallel to the edge (see Figure 4a), the only nonlinear susceptibility component that can therefore contribute to the signal (noting the polarization results in Figure 4) is $\chi_{xxx}$ (as $E_y$~0 and $\vec{E}_{nl}$ can have no $z$ component). Similarly, for SPPs launched by perpendicular gratings the only component that can contribute is $\chi_{xyy}$.

To support our analytical results we used COMSOL to model a single SPP and two counter-propagating SPPs travelling at a gold/air interface, with results presented in Figure S6. As expected, for a single SPP we see extremely weak coupling to far-field SHG, primarily arising from the simulation edges (i.e. the gold/matched layer boundary). Conversely, for two counter-propagating SPPs we observe strong coupling to far-field SHG arising from the entire gold surface, and that this radiation's k-vector is perpendicular to the gold surface.

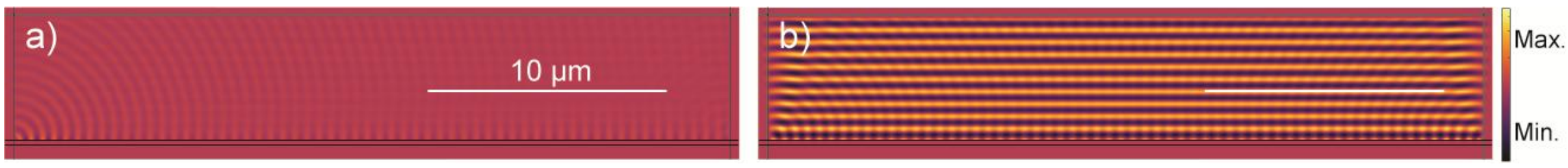

*Figure S6. Second-harmonic electric field component parallel to gold surface arising from a) a single SPP at the gold-air interface (travelling from left to right), and b) counter-propagating SPPs. This electric field component was plotted to study electric field propagating away from the gold surface. Outer boundaries in all simulations are perfectly matched layers. Colour scale is the same for both plots.*

*Supporting Information Note 7 – oscillations observed from monocrystalline surfaces*

We regularly observed oscillations in the SHG signal from flat surfaces. This was always a small component of the total signal, typically less than 20 %. To investigate this further we present the back-focal-plane measurements shown in Figure S5c on a log scale in Figure S7a. Small SHG components are present at higher angles (similar to that observed in other SHG experiments [5]).

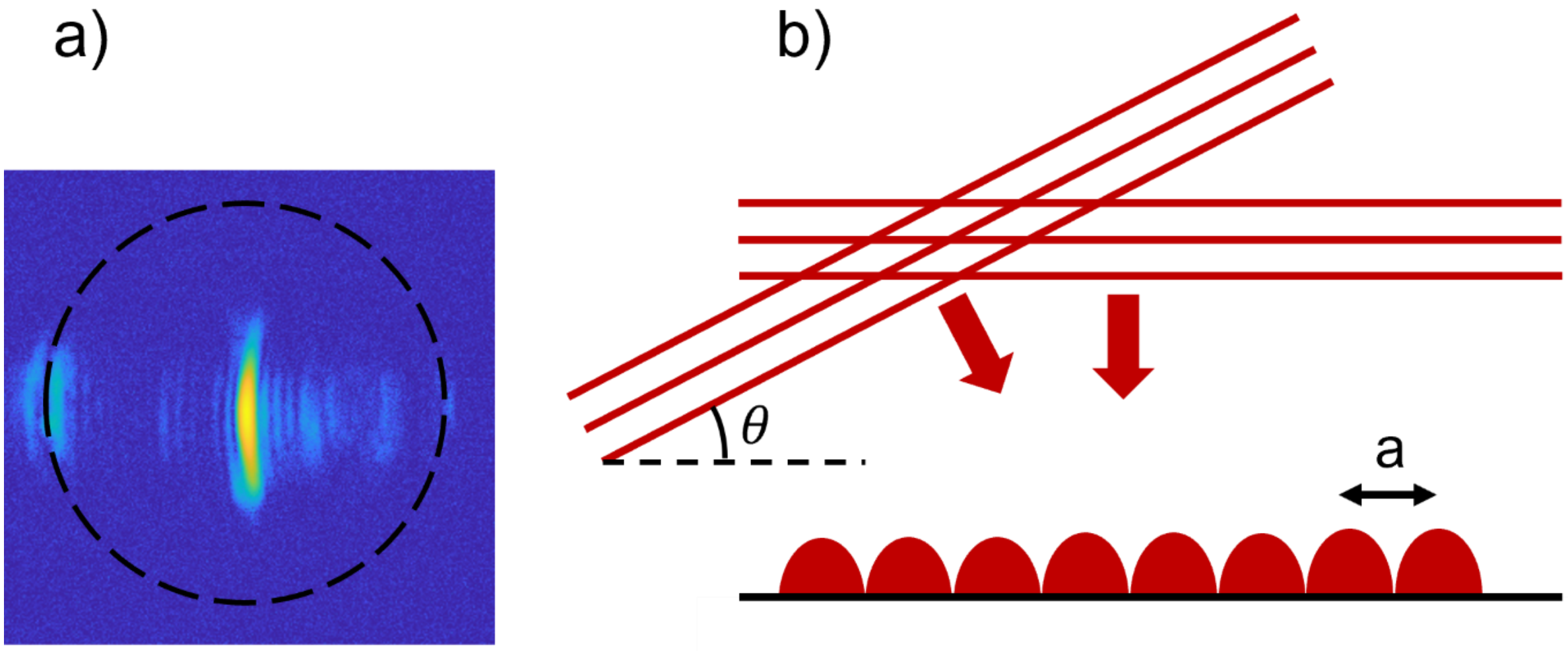

*Figure S7. a) Equivalent of Figure S5c, but on a logarithmic scale. Dashed black line indicates approximate maximum extent of the back-focal plane. b) Schematic of the interference of two plane waves incident on a detector, with one plane wave incident at an angle θ to the detector plane.*

We recorded the spacing of the real space oscillations, which did not correspond to the wavelength of the SPP nor a frequency doubled SPP. Rather, we find that the spacing of the fringes is 0.78 µm for NA of 0.7, but changes with objective NA (0.64 µm/1.13 µm for NA of 0.75/0.45). We consider two plane waves incident on our detector, one perpendicular to the detector and one incident at a high angle, as we recorded in Figure S7a (see schematic, Figure S7b). These plane waves will introduce an interference pattern on the detector. The measured spacing of the fringes (i.e. $a$ on Figure S7b) matched that observed when the high angle wave is incident with the maximum angle (NA) allowed by our objective, confirming that the real-space oscillations are due to the interaction of these two plane waves. This is further demonstrated in Figure S8**:** here we close an iris in the back-focal-plane and record the real-space image. As the range of incident angles is reduced, the spacing of the oscillations reduces i.e. there is a small background SHG component at several angles, and the highest angle component produces the oscillations that we observe. We also note that our back-focal-plane results are fully consistent with taking a Fourier transform of the real space images.

*Figure S8. White light and second-harmonic back-focal-plane, and second-harmonic real space, from a region between two gratings, as an iris is closed in the back-focal-plane. Dashed orange line indicates approximate maximum extent of the back-focal plane.*

The question is therefore what introduces small SHG components at high angles. We carried out a wide range of simulations of SPP induced SHG, including time resolved simulations of this process, but never found these to produce a signal at high angles. Therefore, we considered other processes that could contribute to this signal. Specifically, we note that SHG stronger than that from the SPPs is produced at the gratings (at either end of the SPP induced SHG) and, furthermore, the oscillations are largest closest to these gratings (e.g. Figure 2d, i). SHG from the gratings, unlike that from the SPP induced SHG, is emitted over a wide range of angles (Figure 3b). We now introduce a simplified model of how SHG signal produced over a wide angular range is detected. Specifically, we consider a 1D Gaussian source in the measurement plane, described by:

$$E(x) = e^{-ax^2}. \tag{10}$$

Here $x$ is position along the surface and $a$ describes the spread of the Gaussian. We define the Fourier transform as

$$\tilde{E}(k_x) = \frac{1}{\sqrt{2\pi}} \int e^{-ik_x x} E(x) dx \tag{11}$$

where $k_x$ is the in-plane wavevector. For the Gaussian source its Fourier transform is

$$\tilde{E}(k_x) = \frac{e^{-\frac{k_x^2}{4a}}}{\sqrt{2a}}. \tag{12}$$

In the experiment $k_x$ is only detected up to so some maximum value that is defined by the numerical aperture, $k_{x,0}$. We can define a top hat function such that

$$\Pi_{\mathrm{k}}\left(\frac{k_x}{k_{x,0}}\right) = \begin{cases} 1 \text{ for } |k_x| < k_{x,0} \\ 0 \text{ otherwise} \end{cases}. \tag{13}$$

Therefore, the signal that would be detected with the camera from this Gaussian is given by

$$\tilde{E}_{\mathrm{sig}}(k_x) = \tilde{E}(k_x)\Pi_{\mathrm{k}}\left(\frac{k_x}{k_{x,0}}\right), \tag{14}$$

and in real space this is

$$E_{\mathrm{sig}}(x) = \frac{1}{\sqrt{2\pi}}\int e^{ik_x x}\,\tilde{E}_{\mathrm{sig}}(k_x)dk_x. \tag{15}$$

We can state that $k_{x,0} = \frac{2\pi \mathrm{NA}}{\lambda}$ where NA is the objective numerical aperture and $\lambda$ the measurement wavelength. In our experiment NA=0.7 and $\lambda$ =0.515 μm. As an example, we simulate an initial gaussian which goes to $\frac{1}{e}$ of its maximum intensity over 0.5 μm (i.e. $a = 4$). In Figure S9a we present the simulated Gaussian and the Gaussian as if it were recorded with a microscope with NA=0.7 on a log scale. It can be seen that the limited numerical aperture of the objective introduces small oscillations away from the main peak.

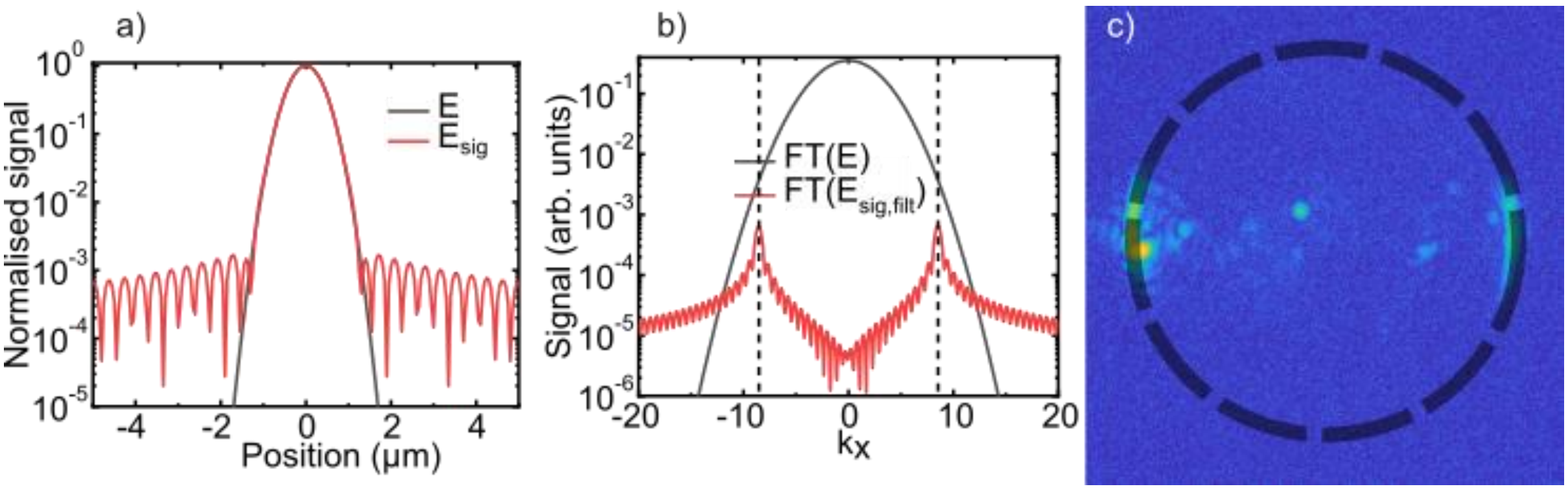

*Figure S9. a) Simulated 1-D Gaussian, and calculated signal when passed through a filter limiting the maximum observable k-space component (see accompanying discussion). b) Fourier transform of simulated full Gaussian signal and of the observable Gaussian, when spatial filtering has been applied in a region away from the main Gaussian peak. Here FT denotes Fourier Transform. Dashed line marks* $k_{x,0}$. *c) Experimental back-focal-plane of region ~10 μm from a grating, with no SHG SPP present,*

*when the grating is excited and produces SHG. Dashed black line represents maximum extent of back-focal-plane.*

We now consider how the oscillations away from the peak appear in k-space (i.e. in a back-focal-plane measurement, like that presented in Figure S7a). To do this we apply spatial filtering to a region away from the Gaussian peak, giving

$$E_{\mathrm{sig,filt}}(x) = E_{\mathrm{sig}}(x)\Pi_{\mathrm{x}}\left(\frac{x - x_1}{x_0}\right) \tag{16}$$

where $x_1$ is the centre of the spatial filter, $2x_0$ the range of the spatial filtering and $\Pi_{\mathrm{x}}$ is a spatial top hat function (rather than in k-space). We can then Fourier transform again to find the k-space components (i.e. back-focal-plane signal):

$$\tilde{E}_{\mathrm{sig,filt}}(k_x) = \frac{1}{\sqrt{2\pi}}\int e^{-ik_x x} E_{\mathrm{sig,filt}}(x)dx\,. \tag{17}$$

This gives us the contribution to the back-focal-plane signal from a Gaussian in a region far from the Gaussian peak. The formula can be re-arranged (via applying the convolution theorem) to give

$$\tilde{E}_{\mathrm{sig,filt}}(k_x) = \frac{x_0 e^{-ix_1 k_x}}{\sqrt{2a}}\int_{-k_{x,0}}^{k_{x,0}} e^{ixk'_x - \frac{k'^2_x}{4a}} sinc\big(x_0(k'_x - k_x)\big)dk'_x\,. \tag{18}$$

To continue our example, we consider spatially filtering the beam 10 μm away from the centre of the Gaussian and apply filtering over a 10 μm region (i.e. $x_0 = 5$, $x_1 = 10$). We present simulations of $\tilde{E}_{\mathrm{sig,filt}}$ in Figure S9b. It can be seen that the main signal contributions are at the objective's limiting numerical aperture.

In summary, the above shows that using a finite numerical aperture objective to observe a signal that has broadband k-space contributions gives peaks in k-space, or back-focal-plane measurements, at the numerical aperture limit set by the objective lens when observing the signal away from its main peak. We propose that this is the situation in our experiments i.e. SHG is produced at a wide range of angles from the gratings. When we observe a region away from the grating peak, the grating gives small contributions to the signal at angles close to the maximum observed value of k-space. The interference of this signal with the SPP induced SHG results in the oscillations observed in our measurement (following the schematic in Figure S7b).

To further support this conclusion we recorded a back-focal-plane signal from a region near an in-coupling grating (i.e. spatially filtering to remove the signal from the grating region), without any reflection grating or other signal present. We again observe strong signal contributions at the maximum spatial extent of the back-focal-plane, as is presented in Figure S9c. This confirms our suggestion that the high angle signal is caused by the gratings rather than the SPP-induced SHG.

*Supporting Information Note 8 – scanning electron microscope images of the fabricated samples*

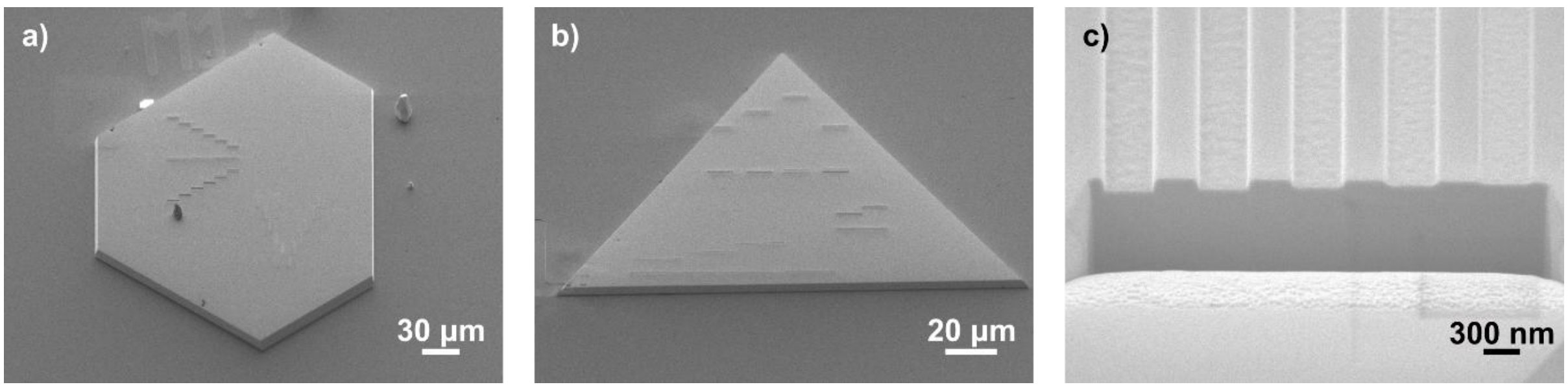

*Figure S10. SEM images of the fabricated samples. a) and b) overview of the various FIB-milled gratings. c) close-up cross-sectional image of an in-coupling grating, showing the depth of the grating grooves and revealing nearly atomic smoothness of unexposed crystalline fold surfaces and high-quality of fabrication.*

*Supporting Information Note 9 – schematic diagram of the nonlinear microscopy setup*

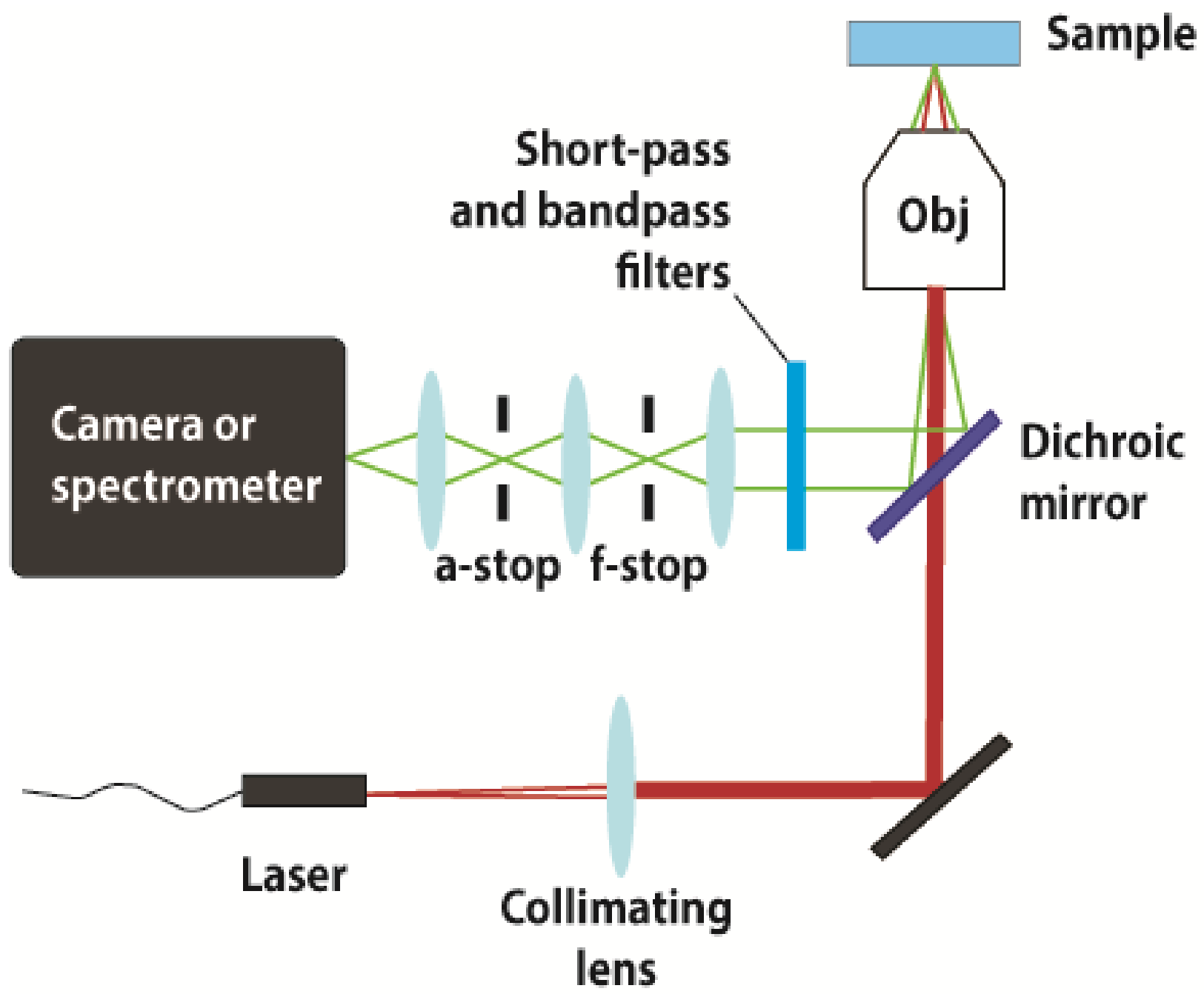

*Figure S11. Schematic of light path used in measurement.*